\documentstyle[twocolumn,aps,psfig]{revtex}
\begin{document}
\draft

%
\twocolumn[\hsize\textwidth\columnwidth\hsize\csname@twocolumnfalse\endcsname
%
%

\title{Study of Spin and Charge Fluctuations in the U-t-${\rm t'}$ Model}

\author{Charles Buhler and Adriana Moreo}

\address{Department of Physics, National High Magnetic Field Lab and
MARTECH, Florida State University, Tallahassee, FL 32306, USA}

\date{\today}
\maketitle

\begin{abstract}
As additional neutron scattering experiments are performed on a variety of high
temperature superconducting compounds it appears that magnetic incommensuration
is a phenomenon common to all of the samples studied. The newest
experimental results indicate that incommensurate peaks exist at
momentum ${\bf q}=\pi(1,1\pm\delta)$ and $\pi(1\pm\delta,1)$ in 
${\rm La_{2-x}Sr_xCuO_4}$ (LSCO) and ${\rm YBa_2Cu_3O_{7-\delta}}$ (YBCO). 
The dependence of $\delta$ with hole doping appears to be similar 
in both materials. In addition,
new ARPES data for LSCO as a function of doping show
that its Fermi surface is qualitatively similar to the one of
YBCO, contrary to what was previously believed.
Early theoretical attempts to explain LSCO and YBCO behavior 
usually relied on one- or three-band Hubbard models or the t-J model
with electron hopping beyond nearest-neighbor and with different parameter
values for each material.
In this paper it is shown that using a one band Hubbard U-t-${\rm t'}$ 
model with a unique set of parameters, U/t=6 and
${\rm t'}$/t=-0.25, good agreement is obtained between computational
calculations and neutron scattering
and ARPES experiments for LSCO and YBCO. It is also shown that using
a more negative ${\rm t'}$/t will induce short-range magnetic 
incommensuration along the diagonal
direction in the Brillouin zone, in qualitative
disagreement with the experimental results. At the finite temperatures
of the present Monte Carlo simulation it is also observed that in this
model the tendency to 
incommensuration appears to be more related to the shape of the 
two-dimensional Fermi surface and the strength of the interaction, rather
than to charge order.

\end{abstract}

\vskip2pc]
\narrowtext

\section{Introduction}

Neutron scattering experiments continue providing exciting information about
the behavior of spin and charge degrees of freedom in the high $T_c$
cuprates. Studies on ${\rm La_{2-x}Sr_xCuO_4}$ (LSCO)\cite{Cheong} have 
shown the
existence of incommensurability in the spin channel near the
commensurate position $(\pi,\pi)$. In this case, the incommensurate
peaks were found at $Q_{\delta}=\pi(1\pm\delta,1)$ and
$\pi(1,1\pm\delta)$. The value of $\delta$ increases linearly with
doping in the range ${\rm0.05\leq x\leq 0.14}$, while 
for ${\rm x\geq 0.14}$, $\delta$ plateaus
at 0.25.\cite{yamada} The
intensity  at $Q_{\gamma}=\pi(1\pm\delta/2,1\pm\delta/2)$
was observed to be 0.18 times that at $Q_{\delta}$ for x=0.14 while the
intensity at $Q_{\pi}=(\pi,\pi)$ was negligible compared with the
background.\cite{Mason}
For some time neutron
scattering experiments in ${\rm YBa_2Cu_3O_{7-\delta}}$ (YBCO) 
were less clear. Rossat-Mignod and coworkers detected
magnetic fluctuations only at the commensurate position\cite{Ros}, while
Tranquada et.al\cite{Tran1} noticed possible incommensurate
fluctuations. Measurements by Dai et al.\cite{Dai} using a new
position-sensitive detector bank
indicate that incommensurability in the spin channel indeed is present
in YBCO. Although originally incommensurability was detected only along the
diagonal direction in the Brillouin zone, the latest results indicate
that it occurs at $Q_{\delta}$ as in the case of LSCO, and the
dependence of $\delta$ with doping appears to be similar in both 
materials.\cite{Mook1} 
In addition, ${\rm Bi_2Sr_2CaCu_2O_8}$
has been recently studied by Mook and Chakoumakos\cite{Mook} and an
incommensurate fluctuation that occurs below $T_c$ was found.
This
incommensuration 
was identified with a dynamic charge density wave because its scattering
intensity appeared to increase with increasing momentum transfer.
Scattering that could be described as magnetic has not been observed yet
in this material because the experimental technique used
does not allow to reach high enough values of the momentum transfer.
However, magnetic excitations should be present if a dynamic strip 
phase gives origin to the charge peaks. 
Evidence of incommensurability in the charge channel
was also observed in ${\rm La_{1.6-x}Nd_{0.4}Sr_xCuO_4}$\cite{Tran} which
led to speculations about a similar behavior in LSCO. This would
indicate that the incommensuration
in the spin channel may be due to the existence of charge-stripe order
rather than to some kind of charge uniform spiral spin state or 2D Fermi
surface effects.\cite{Little} 

While some of these experiments suggest that magnetic
incommensurability in the cuprates maybe due to charge-stripe order in
which the orientation of the stripes is not material dependent, 
the theoretical understanding of this phenomenon is
less clear. Short-range magnetic incommensurate correlations in 
the spin channel were detected
early on in the Hubbard\cite{Adri,Ima} and t-J\cite{Adri1} models. The split
of the commensurate peak was observed to be qualitatively similar to the
behavior in LSCO and YBCO. However, the dependence of $\delta$ with
doping did not reproduce the experimental data and, as it will be shown
later, the experimentally observed 
relative intensities of the peaks at different points in the
Brillouin zone are not well reproduced either.
No particular order was reported in the charge
channel in these models through numerical analysis.\cite{Chen,Put} 
However, since the proposal of dynamical microphase
separation in the ${\rm CuO_2}$ planes\cite{Eme}, a reanalysis of these
conclusions is needed.

At the same time that the recent neutron scattering experiments were 
discovering new similarities between the magnetic properties of the 
different cuprates, also recent
angular-resolved photoemission (ARPES) experiments performed on LSCO
mapped its Fermi surface (FS)\cite{Fuji} at different values of doping
unveiling interesting similarities 
among the qualitative FS shape of several high $T_c$
materials. 

Motivated by the new experimental results the aim of this paper is to 
revisit the U-t-${\rm t'}$ model, exploring numerically the spin and charge
channels to determine whether it is possible to obtain agreement
with the new data for LSCO and YBCO using a unique set of parameters in
the model. It will be concluded that this is indeed possible. The paper is
organized as follows: in Section II the ${\rm U-t-t'}$ Hamiltonian and
the notation used are introduced. Results on magnetic and charge
correlations, as well as on the shape of the Fermi surface, are presented
in Section III while Section IV is devoted to the conclusions.

\section{The Model}

The U-t-${\rm t'}$ one band Hubbard model Hamiltonian is given by

$$
{\rm H=
-t\sum_{<{\bf{ij}}>,\sigma}(c^{\dagger}_{{\bf{i}},\sigma}
c_{{\bf{j}},\sigma}+h.c.)
-t'\sum_{<{\bf{in}}>,\sigma}(c^{\dagger}_{{\bf{i}},\sigma}
c_{{\bf{n}},\sigma}+h.c.)+}
$$

$$
{\rm U\sum_{{\bf{i}}}(n_{{\bf{i}} \uparrow}-1/2)( n_{{\bf{i}}
\downarrow}-1/2)+\mu\sum_{{\bf{i}},\sigma}n_{{\bf{i}}\sigma} },
\eqno(1)
$$
\noindent where ${\rm c^{\dagger}_{{\bf{i}},\sigma} }$ creates an electron at
site ${\rm {\bf i } }$
with spin projection $\sigma$, ${\rm n_{{\bf{i}}\sigma} }$ is the number
operator, the sum
${\rm \langle {\bf{ij}} \rangle }$ 
runs over pairs of  nearest neighbor lattice sites, and the sum
${\rm \langle {\bf{in}} \rangle }$ 
runs over pairs of lattice sites along the plaquette
diagonals. ${\rm U}$ is the
on site Coulombic repulsion, ${\rm t}$ the nearest neighbor hopping
amplitude, ${\rm t'}$ the diagonal hopping amplitude, and $\mu$ the
chemical potential. In this work ${\rm t}$ will be set equal to 1.

The static charge and magnetic structure factors ${\rm N({\bf q})}$ and
${\rm S({\bf q})}$ are defined by the relations

$$
{\rm N({\bf q})=\sum_{\bf r} e^{i{{\bf q} \cdot{\bf r}}} \langle 
\delta n_{\bf 0} \delta n_{\bf r} \rangle},
\eqno(2)
$$

$$
{\rm S({\bf q})=\sum_{\bf r} e^{i{{\bf q} \cdot{\bf r}}} \langle S^z_{\bf 0} S^z_{\bf r} \rangle},
\eqno(3)
$$

\noindent where $\langle\delta n_{\bf 0}\delta n_{\bf r}\rangle$ and 
$\langle S^z_{\bf 0}S^z_{\bf r} \rangle$ 
are equal-time density- and spin-correlation
functions,   
${\rm S^z_{\bf
r}={1\over{2}}\sum_{\alpha,\beta}c^{\dagger}_{{\bf r},\alpha}
\sigma^z_{\alpha,\beta}c_{{\bf r},\beta}}$, and ${\rm \delta n_{\bf
r}=\sum_{\sigma} c^{\dagger}_{{\bf r},\sigma}c_{{\bf
r},\sigma}-\langle n \rangle}$. Here $\rm{\langle n \rangle=1-x}$
is the average density of electrons. The brackets in Eqs. (2) and (3)
refer to thermal averaging in the grand canonical ensemble which will be
performed using the standard Quantum Monte Carlo (QMC) determinantal method.

Before presenting our results let us discuss the behavior of the spin
and charge correlations in the non-interacting system (U/t=0), 
and also in the standard Hubbard
model with ${\rm t'}$=0. In the non-interacting system the spin and charge
correlations are related through ${\rm S({\bf q})={1\over{4}}N({\bf
q})}$. ${\rm N({\bf q})}$ increases from zero reaching the value
$\langle n \rangle$ at ${\rm q=2k_F}$ and remaining constant
afterwards. On the other hand, 
in the Hubbard model it was observed that at low density, i.e.,
 $\langle n \rangle< 0.5$, ${\rm S({\bf q})}$ peaks at ${\rm
q=2k_F}$ while ${\rm N({\bf q})}$ is suppressed at these momenta
compared with the non-interacting system and it only peaks at 
${\bf q}=(\pi,\pi)$.\cite{Chen} This peak is due to the short range
effective repulsion between particles. The behavior at higher densities
is very different in the spin channel. At half-filling a sharp peak
develops at $Q_{\pi}$ and, with a substantially reduced intensity, it moves to
$Q_{\delta}$ with doping. While the intensity is maximum at $Q_{\delta}$
as in the experiments, the weight at $Q_{\pi}$ is always larger than
the one at $Q_{\gamma}$ \cite{Adri} 
which is qualitatively incorrect compared with recent experiments 
as it will be shown below.

\section{Results}

Due to the well known ``sign problem'' it is very difficult to perform
Monte Carlo numerical studies at small hole 
doping, low temperatures and values of ${\rm U/t
>4}$. This problem is exacerbated as the absolute value of ${\rm t'}$
increases. For this reason the numerical
efforts will be concentrated here on the study of the fixed 
density $\langle n \rangle=0.7$, (i.e.,
x=0.3) since for this $\langle n \rangle$ a good degree of control of the
numerical results can be achieved.
In addition, new experiments have been
performed in LSCO at precisely x=0.3 providing information about
incommensuration \cite{yamada} and the shape of the FS\cite{Fuji}.
Here  results for U/t=6 on $8 \times 8$ lattices will be
presented. Due to
the sign problem at this relatively large value of U/t, and using a
finite diagonal hopping ${\rm t'}$, the temperature had to be fixed at
T=0.25t. It is to be expected that the intensity of the observed
features will increase at lower temperatures and the results presented
here are, thus, lower bounds to the actual values. 

Note that, as part of our study, runs for other values of U/t were 
performed and qualitative differences with the results for U/t=6 were 
not observed. In particular, for ${\rm t'/t=0}$ we
found that for values of U/t as high as 10 the magnetic
incommensuration always occurs at $Q_{\delta}$ rather than along the
diagonal as predicted by mean-field calculations in the strong coupling
regime.\cite{Chub}

\subsection{Magnetic Incommensurability}

As a first step, the static structure factor will be calculated for
several values of ${\rm t'/t}$ and comparisons with the experimental
neutron scattering results will be made along several directions in 
momentum space.

In Fig.1-a the peaks in $S({\bf q})$, indicative of short range spin
incommensurate tendencies in the ${\rm U-t-t'}$ model, are presented along the
$(0,\pi)-(2 \pi,\pi)$ direction for values of ${\rm t'}$/t ranging from
0 to -0.5. The figure shows
that for all values of ${\rm t'/t}$ analyzed here, the peak in the structure
factor occurs at ${\bf q}=(3 \pi/4,\pi)$ and $(5 \pi/4,\pi)$ 
which correspond to $\delta=0.25$. This is in agreement with
the experimental value for x=0.3 in LSCO. \cite{yamada} However, a spline fit
through the available 
data points (dashed line) suggests that the actual peak at
${\rm |t'/t|\geq 0.3}$ occurs at a slightly larger value of $\delta$. 
Actually, when
the structure factor is scanned along the diagonal direction as shown in
Fig.1-b it is clear that the results with ${\rm |t'/t|=0.3}$ or larger do not
fit the experimental data because the maxima in $S({\bf q})$ at ${\bf
q}=(1\pm\delta)\pi (1,1)=(3 \pi/4,3 \pi/4)$ and $(5\pi/4,5\pi/4)$  
have intensities
which are approximately equal or higher than those
at $Q_{\delta}$\cite{Man}, as it can be deduced by comparing Fig.1-a
with Fig.1-b. This is in disagreement with the 
experiments that indicate that the intensity at ${\bf
q}=(1\pm\delta)\pi (1,1)$ should be indistinguishable from the 
background.\cite{yamada,Mason,Dai,Mook1}

An important result is that at ${\rm t'}$=0 $S({\bf q})$ 
along the diagonal direction, ${\bf q_x=q_y}$,
has a maximum at ${\bf q}=Q_{\pi}$ as it can be observed 
in Fig.1-b (top curve). 
This behavior is in disagreement with the
experimental data for LSCO presented in Fig.3 (closed circles) of Ref. 
\cite{Mason}, and for YBCO in Fig.1-c of Ref. \cite{Dai} where a
minimum is observed at $Q_{\pi}$ along the diagonal. The experimental
results were obtained at x=0.15 for optimally doped LSCO and 
x=0.1 (i.e., $\delta=0.4$) for YBCO showing that
the qualitative behavior does not depend strongly on doping. 
This indicates that the standard
Hubbard model (${\rm t'=0}$) 
does not describe the qualitative behavior observed with
neutron scattering when $\delta=0.25$. However,
working with ${\rm t'}$=-0.2 the fit along the diagonal in momentum
space (Fig.1-b)
indicates that $S(Q_{\pi})$ now has become a local minimum, in qualitative
agreement with the experiments. Then, the constraints on the relative
intensity of the peaks provided by the experiments leave
a small window of possible values of ${\rm t'}$/t. If ${\rm |t'/t|\ge
0.3 }$ the incommensurate peaks will appear along the diagonal rather
than at $Q_{\delta}$ (as discussed in the previous paragraph) 
and if ${\rm |t'/t|\le 0.2}$ the structure factor
at $Q_{\pi}$ would be a local maximum rather than a minimum along the
diagonal direction.

Additional experimental measurements\cite{Mason} 
have been reported along the direction
$Q_{\delta}[=\pi(1-\delta,1)]\rightarrow Q_{\gamma}[=\pi(1-\delta/2,
1+\delta/2)]\rightarrow Q_{\delta}[=\pi(1,1+\delta)]$.
Note that for
$\delta=0.25$ the point $Q_{\gamma}$ is given by $(7\pi/8,9\pi/8)$ which
is computationally accessible only on $16 \times 16$ or
larger lattices if periodic boundary conditions are being used. Since
such a large lattice size is beyond the capabilities of the QMC at
finite hole density we obtained the values of $S({\bf q})$ at the equivalent
$Q_{\gamma}$ points located at $7\pi/8(1,1)$ and
$9\pi/8(1,1)$ with the help of a spline fit of the available data in Fig.1-b 
(dashed line).

\begin{figure}[htbp]
\centerline{\psfig{figure=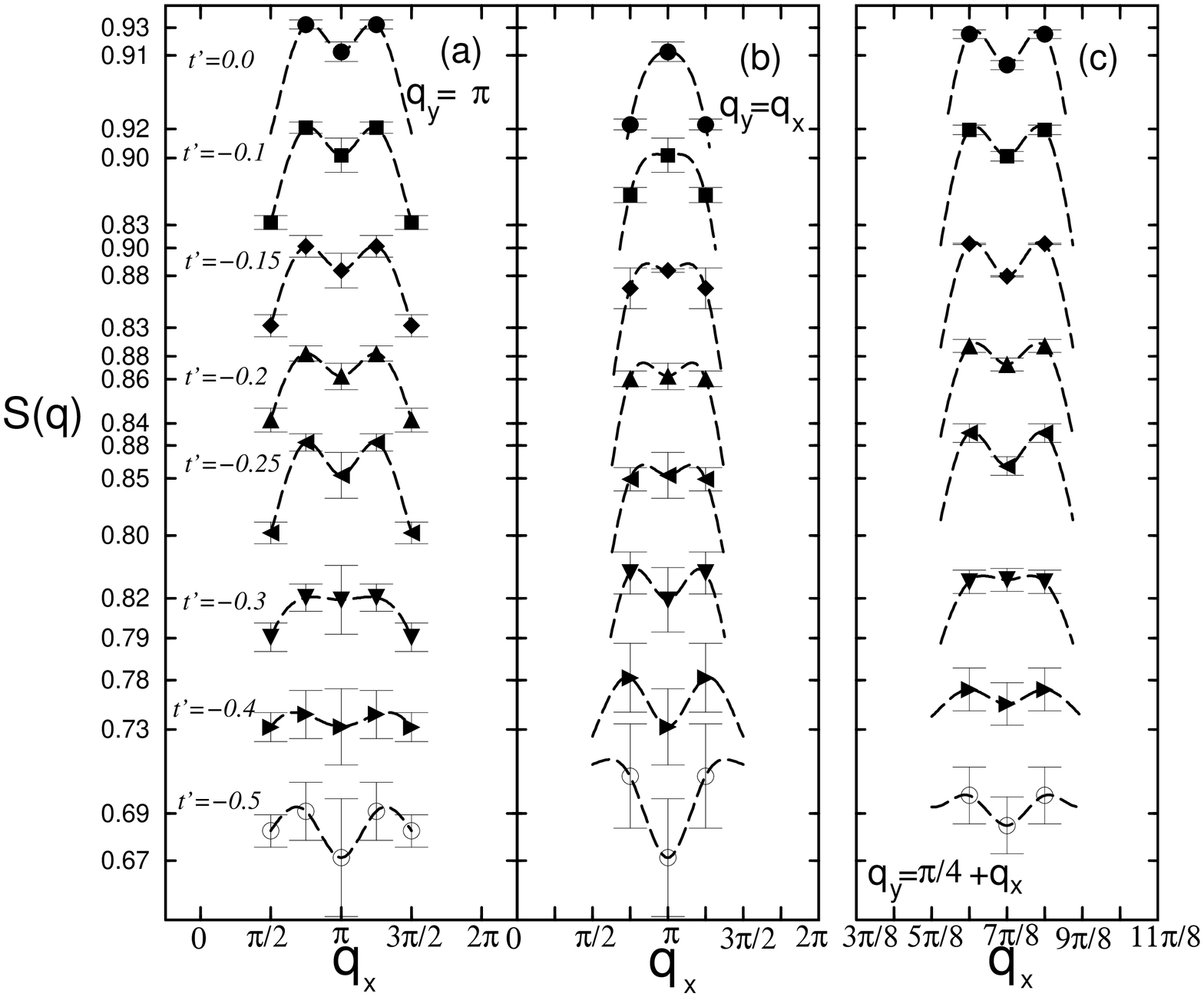,width=8.5cm,angle=-90}}
\vspace{0.3cm}
\caption{The static structure factor $S({\bf q})$ for U/t=6, $\langle n
\rangle=0.7$ on an $8 \times 8$ lattice for T=0.25t and values of
${\rm t'}$/t ranging from 0 to -0.5. a) Along the 
$(\pi,0)-(\pi,2\pi)$ direction; b) along the diagonal direction; 
c) along the $q_y = \delta \pi + q_x$
direction. The dashed line indicates a spline
fit.}

\vspace{0.5cm}
In Fig.1-c the structure factor is shown along the
line  $q_y = \delta \pi + q_x$ which corresponds to the direction along
which the experimental neutron scattering 
data shown by open circles in Fig.3 of Ref. \cite{Mason} were
taken. In the figure it can be seen that the relative intensity of the
numerical data at
$q_x=\pi(1-\delta/2)=7\pi/8$, (i.e., $Q_{\gamma}$) and $q_x=\pi (1\pm
\delta)=0.75\pi$ or $1.25\pi$,
(i.e., $Q_{\delta}$) is a function of ${\rm t'}$. Considering, as in the
experiments,  that the
intensity for ${\bf q}=\pi (1\pm \delta) (1,1)$ (this point corresponds
to $q_x=3\pi/4$ and $5 \pi/4$ in Fig.1-b) has to correspond to the
background, then the relationship
$Q_{\gamma}/Q_{\delta}=0.18$ is satisfied for ${\rm t'}$/t=-0.25. This is used
as a guide because this relationship may be doping dependent but since
the window in ${\rm t'}$/t is so narrow, at most an error of the order of 0.05
in the estimation of ${\rm t'}$/t is being made.
 
The above analysis shows that a
comparison of the present numerical data when the measured incommensurability
$\delta$ is 0.25 agrees with the data for $LSCO$ with the same $\delta$
using U/t=6 and ${\rm t'}$/t=-0.25. Notice that the position and the relative
intensity of the peaks does not change too much with temperature
according to the experiment. 

\subsection{Fermi Surface}

The next issue that will be addressed is whether the ratio  ${\rm
t'/t=-0.25}$, fixed by the spin structure factor 
analysis in the previous subsection,
will fit other experimental data such as, for example, the shape of the
Fermi surface recently obtained using ARPES for LSCO.\cite{Fuji} 
The possible shape of the FS will be determined by analyzing the
momentum distribution ${\rm n({\bf q})}$ which is calculated by Fourier
transforming the one-electron Green's function,

$$
{\rm g_{\bf ij}=-\langle\sum_{\sigma} c_{{\bf i},\sigma}
c^{\dagger}_{{\bf j},\sigma}\rangle},
\eqno(4)
$$

\noindent that is evaluated using QMC.

The
criteria used here to obtain the most probable locus of the FS from
numerical ${\rm n({\bf q})}$ data are two: (a) find the values of ${\bf q}$
where ${\rm n({\bf q})}\approx 0.5$\cite{Adri} and (b) find the values of ${\bf q}$
where ${\rm n({\bf q})}$ changes the most rapidly\cite{Dan}. For the case of
${\rm t'}$/t=-0.25 we have 
observed that both methods provide similar results in the
regions close to the diagonal direction in the Brillouin zone, but substantial
differences were observed close to the $(0,\pi)$ and $(\pi,0)$ points.
While criterion (a) indicated a FS closed around (0,0), criterion (b) indicated
a FS closed around $(\pi,\pi)$. Since criterion (b) provided
similar results for the non-interacting case  with $\langle n
\rangle=0.7$ and ${\rm t'}$/t=-0.25, a situation where it is known
that the FS actually closes around
(0,0), it was decided that criterion (a) would be more effective in this
context. 

The FS obtained with procedure (a) is shown in Fig.2-a. The closed
circles indicate where ${\rm n({\bf q})}\approx 0.5$ and the continuous line
is a 6th order polynomial fit of the points. The obtained FS is very
similar to the
non-interacting one (denoted by a dashed line in the same figure). 
It satisfies Luttinger's theorem within error bars
and, actually it is in excellent agreement with the experimental data for
x=0.3\cite{Fuji} shown with open squares in Fig.2-a. Thus, using ${\rm
t'/t=-0.25 }$ good agreement
has been obtained between the numerical results and two independent
experiments (neutron scattering and ARPES) performed in overdoped LSCO.
Since the obtained Fermi surface resembles closely the non-interacting one, 
the FS in the U/t=0 limit was calculated for x=0.1 and its shape compared
directly with the experimental 
results at this density. As it can be observed in Fig.2-b the
agreement is once 
again very good. According to Ref. \cite{Fuji} the FS for LSCO at
optimal doping is still centered about $(\pi,\pi)$ and it is
qualitatively similar to the one obtained for x=0.1. This is indeed what
happens with the U/t=0 FS for ${\rm t'}$/t=-0.25 at x=0.15 which is shown 
in Fig.2-c (dashed line). 
At the non-interacting level the change between a FS that
closes around $(\pi,\pi)$ and around $(0,0)$ occurs at x=0.22 for the
ratio of ${\rm t'/t}$ used here. In addition,
in Fig.2-c the experimental points for optimal
doped YBCO obtained several years ago\cite{campu} are also shown. 

The agreement with our result is only
qualitative but it has to be considered that 
the measurements are very difficult due to surface effects and thus the
experimental points have
large error bars (not shown). It is important to remark
that the similarity found between the interacting and the
non-interacting FS for ${\rm t'}$/t=-0.25 does not mean that they are
identical but the differences will be apparent only when larger lattices
at lower temperatures can be studied.

Sometimes the experimental results
have been interpreted as indicating that there is a FS only
along the diagonal direction and no FS close to $(0,\pi)$ and 
$(\pi,0)$.\cite{arc}
The rate of change of $n({\bf q})$ could support this view but we found
out that it also gets very reduced close to $(0,\pi)$ and $(\pi,0)$ 
even in the non-interacting
case when it is known that there is a continuous FS. Thus, the present
results do not allow us to decide one way or the other.
We also noticed that the interacting FS for ${\rm t'}$/t=-0.3 appears to be
qualitatively different from the non-interacting one. In particular at
x=0.3 it seems to close around $(\pi,\pi)$ while the non-interacting
one closes around $(0,0)$. Thus, if YBCO were to have the same
qualitative shape of FS than LSCO in the overdoped regime this is
another reason to rule out values of ${\rm |t'/t|}$ equal or higher than 0.3. 

\begin{figure}[htbp]
\centerline{\psfig{figure=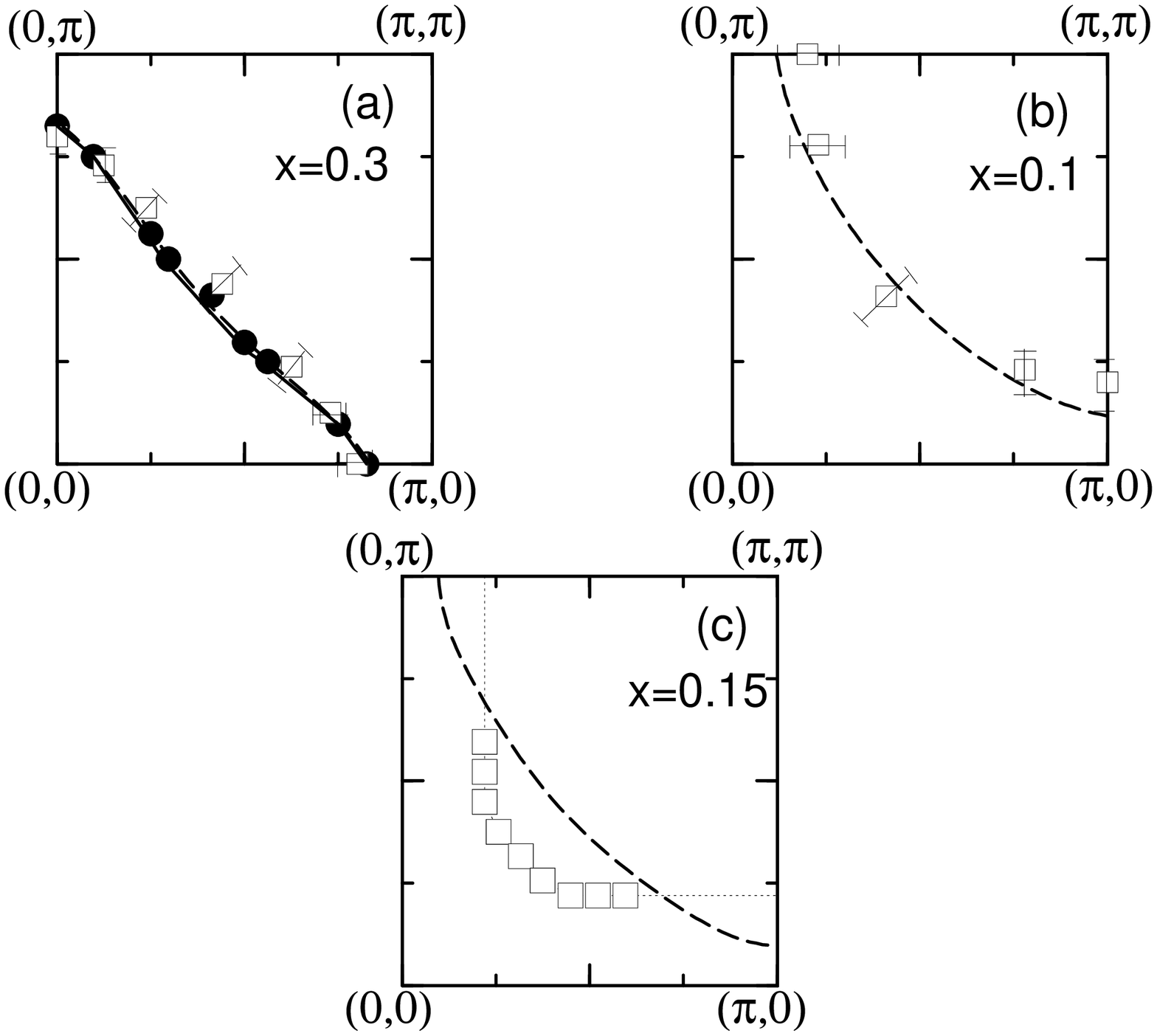,width=8.5cm,angle=-90}}
\vspace{0.5cm}
\caption{ (a) The numerically calculated Fermi surface for U/t=6, $\langle n
\rangle=0.7$ on an $8 \times 8$ lattice for T=0.25t and ${\rm t'}$/t=-0.25
(closed circles and solid line); the open squares are experimental
results for LSCO at x=0.3 from Ref. [17]; the dashed line is the
non-interacting, U/t=0,  FS for ${\rm t'}$/t=-0.25 and $\langle n
\rangle=0.7$; (b) non-interacting (U/t=0) FS for ${\rm t'}$/t=-0.25 and
density 0.9 (dashed line) and experimental data for LSCO with x=0.1 from
Ref. [17]; (c) non-interacting FS for ${\rm t'}$/t=-0.25 and
density 0.85 (dashed line) together with  experimental data for YBCO
with ${\rm x\approx 0.15}$ from Ref. [22].}

\end{figure}

\subsection{Charge Correlations}

The next issue that will be addressed is the origin of the incommensurate
magnetic fluctuations in the present model. In Fig.3 the charge
structure factor $N({\bf q})$ along the
$(0,0)\rightarrow(\pi,\pi)\rightarrow(\pi,0)\rightarrow(0,0)$
directions for ${\rm t'}$/t ranging from 0 (top) to -0.5 (bottom) is shown. 
In all cases there is a
broad maximum at $Q_{\pi}$ which is due, as in the low electron density limit
of the Hubbard model, to the short range effective repulsion between
particles.

If the incommensurate magnetic fluctuations were due to
dynamical charge fluctuations, peaks at ${\bf q}=\pi(0,2\delta)$ and
$\pi(2\delta,0)$ should be observed in $N({\bf q})$ according to
previous theoretical studies\cite{Tran,Eme}. 
In the present case, since $\delta=0.25$, the peaks
would be expected at $(0,\pi/2)$ and $(\pi/2,0)$. This momentum is
indicated with an arrow in Fig.3 and it is clear from the figure that  
no indications of incommensurate charge order is
observed.

Another possible origin of the magnetic incommensuration in 2D could be
simple FS effects. \cite{Little,benard,si} There are some 
momenta that map points (or regions) of the FS into other
points (or regions) also on the FS. These are called nesting or
pseudonesting vectors and they correspond to values of momentum where
maxima occur in the imaginary part of the magnetic susceptibility in the
2D non-interacting system. We performed a numerical calculation of the
pseudonesting vector for the interacting FS.  We constructed a histogram
in order to identify the value of the momentum that mapped most points
of

\begin{figure}[htbp]
\centerline{\psfig{figure=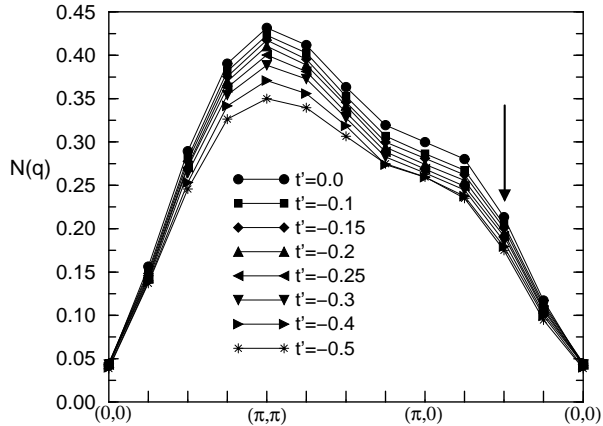,width=8.5cm,angle=-90}}
\caption{Static charge structure factor for U/t=6, $\langle n
\rangle=0.7$ on an $8 \times 8$ lattice for T=0.25t and values of
${\rm t'}$/t ranging from 0 (top) to -0.5 (bottom) along the directions 
$(0,0)-(\pi,\pi)-(\pi,0)-(0,0)$. The arrow indicates the value of the
momentum where a maximum indicating incommensurate short-range order 
would be expected.}
\end{figure}

\noindent{the FS into other points also belonging to it.  For
${\rm t'}$/t=-0.25 at x=0.3 a maximum in the histogram was obtained at ${\bf
q}=(\pi, 0.67 \pi)$ which is in very good agreement with the analytical
value for the corresponding non-interacting FS, namely ${\bf
q}=(\pi, 0.71 \pi)$\cite{benard}. Thus, the maximum in $S({\bf q})$ at ${\bf
q}=(\pi, 0.75 \pi)$ could be explained by FS effects in this case but it
may be due to the coarse grid that necessarily had to be used in our
computational studies. In the
non-interacting case it is expected that the maximum remains at
$Q_{\pi}$ until x reaches 0.22 (see Sect.III.B). 
Though this behavior seems to be in
agreement with previous results for the interacting case \cite{Dan} 
and in disagreement with the experimental data, it is possible that 
the effect of the interaction at smaller dopings will be observed at lower
temperatures than the ones that can presently be reached.}

\section{Conclusions}

In summary, motivated by recent neutron scattering and ARPES
experiments,  we have studied the U-t-${\rm t'}$ model numerically on $8 \times
8$ lattices at temperature T=0.25t and with coupling U/t=6. 
Fixing the density at $\langle n \rangle=0.7$
agreement between the position and the relative intensity of
the incommensurate peaks obtained numerically and  
experimentally for LSCO and YBCO is
observed for ${\rm t'}$/t=-0.25. Larger values of $|{\rm t'/t|}$ 
are ruled out
because in this case the structure factor has a maximum along the
diagonal direction rather than at $Q_{\delta}$, while with less negative
values of ${\rm t'/t}$ 
a relative maximum is observed at $Q_{\pi}$ along the diagonal
direction, again in disagreement with experimental results. Thus, the
Hubbard or t-J models without nearest-neighbor electron
hopping do not reproduce appropriately this experimental 
behavior in the cuprates.

The addition of a diagonal hopping ${\rm t'}$/t=-0.25 also provides 
good agreement with experimental 
measurements of the FS for LSCO and YBCO at different densities. 
A FS that closes around $(\pi,\pi)$ in the underdoped and optimal
doped regimes, and around $(0,0)$ in the overdoped case is observed. 

The
incommensurate magnetic peaks do not seem related to incommensurate charge
fluctuations. In the overdoped regime FS effects enhanced by the
electronic interactions may be responsible for the observed results. The sign
problem prevented the exploration of this issue in the optimal doped
and underdoped regimes. 

The present numerical analysis provides good agreement between a
theoretical model and
two unrelated experiments in the overdoped regime of LSCO. It
also shows that the consideration of a diagonal 
hopping in models
\vspace{1cm}

\noindent{for the cuprates is crucial in order to reproduce 
experimental data. Note that this result is in excellent agreement with
ARPES calculations that have focussed on the insulating compound ${\rm
Sr_2CuO_2Cl_2}$.\cite{tpp} For the cases analyzed here, the same ratio ${\rm
t'/t}$ can reproduce results for both LSCO and YBCO. Although the
behavior of the incommensurate magnetic peaks in the cuprates appear to
be similar, there is still no experimental information about relative
intensities of the peaks at different points in the Brillouin
zone carried out at the same density. These measurements could indicate
possible material dependent properties that could depend on 
longer range electron hopping terms.\cite{tpp}}

\section{Acknowledgements}

We would like to acknowledge useful discussions with P. Dai, G. Aeppli, T.
Mason, H. Mook, Q. Si, J. Tranquada, M. Randeria, M. Norman
and E. Dagotto. A.M. is supported by NSF under grant DMR-95-20776.
Additional support is provided by the National High Magnetic Field Lab 
and MARTECH.
 
\vfil\eject

%
%






\end{figure}
\end{document}